\documentclass[conference]{IEEEtran}
\IEEEoverridecommandlockouts
\usepackage{cite}
\usepackage{amsmath,amssymb,amsfonts}
\usepackage{algorithmic}
\usepackage{graphicx}
\usepackage{textcomp}
\usepackage{xcolor}
\def\BibTeX{{\rm B\kern-.05em{\sc i\kern-.025em b}\kern-.08em
    T\kern-.1667em\lower.7ex\hbox{E}\kern-.125emX}}

\makeatletter
\newcommand{\linebreakand}{%
    \end{@IEEEauthorhalign}
    \hfill\mbox{}\par
    \mbox{}\hfill\begin{@IEEEauthorhalign}
}

\begin{document}
\title{E-commerce Transactions in Islam: Fiqh Muamalah on The Validity of Buying and Selling on Digital Platforms\\}

\author{\IEEEauthorblockN{1\textsuperscript{st} Wisnu Uriawan}
\IEEEauthorblockA{\textit{Informatics Department}\\
\textit{UIN Sunan Gunung Djati Bandung}\\
Jawa Barat, Indonesia\\
wisnu\_u@uinsgd.ac.id}
\and
\IEEEauthorblockN{2\textsuperscript{nd} Muhammad Farhan Tarigan}
\IEEEauthorblockA{\textit{Informatics Department}\\
\textit{UIN Sunan Gunung Djati Bandung}\\
Jawa Barat, Indonesia\\
mhfarhianalmz@gmail.com}
\and
\IEEEauthorblockN{3\textsuperscript{rd} Herdin Kristianjani Zebua}
\IEEEauthorblockA{\textit{Informatics Department}\\
\textit{UIN Sunan Gunung Djati Bandung}\\
Jawa Barat, Indonesia\\
kzherdin03@gmail.com}
\linebreakand
\IEEEauthorblockN{4\textsuperscript{th} Muhamad Nopid Andriansyah}
\IEEEauthorblockA{\textit{Informatics Department}\\
\textit{UIN Sunan Gunung Djati Bandung}\\
Jawa Barat, Indonesia\\
muhamadnopidandriansyah@gmail.com}
\and
\IEEEauthorblockN{5\textsuperscript{th} Marleni Sukarya}
\IEEEauthorblockA{\textit{Informatics Department}\\
\textit{UIN Sunan Gunung Djati Bandung}\\
Jawa Barat, Indonesia\\
marlenijournal@gmail.com}
\and
\IEEEauthorblockN{6\textsuperscript{th} Muhammad Rafli Haikal}
\IEEEauthorblockA{\textit{Informatics Department}\\
\textit{UIN Sunan Gunung Djati Bandung}\\
Jawa Barat, Indonesia\\
muhammadraflihaikal2004@gmail.com}
}

\maketitle

\begin{abstract}
The development of the digital economy has established \textit{e-commerce} platforms as the primary space for commercial transactions for the Muslim community. However, innovations in features and business models on these platforms have gave rise to Sharia issues that cannot be fully explained through conventional Fiqh Muamalah contract frameworks. This research aims to examine the compliance of transaction practices in \textit{e-commerce} with Sharia principles, particularly in the six most frequently used transaction forms, namely information arbitrage-based \textit{dropshipping}, \textit{Buy Now Pay Later} (BNPL) financing schemes, digital representations potentially containing \textit{tadlis} and \textit{ghabn}, algorithmic marketing that encourages consumptive behavior, product halal verification, and \textit{Pre-Order} (PO) transaction models. The research method used is a \textit{Critical Literature Review} with a normative juridical approach, through the study of arguments from the Qur'an, Hadith, DSN-MUI Fatwas, as well as classical and contemporary fiqh literature. The results show that \textit{dropshipping} and PO practices are considered invalid if conducted with a direct sale contract (\textit{bai'}) due to the non-fulfillment of the element of possession (\textit{qabd}) and the presence of high uncertainty (\textit{gharar}). Both practices can be justified through the restructuring of contracts into \textit{wakālah bil ujrah}, \textit{salam}, or \textit{istishna'}. Conventional BNPL is declared non-compliant with Sharia because it contains \textit{riba nasiah} and \textit{riba qardh}. Misleading digital representations and halal claims without valid verification fall into the category of \textit{tadlis}, while \textit{dark patterns}-based algorithmic marketing contradicts \textit{maqāshid al-syariah}, especially the protection of wealth (\textit{hifz al-māl}) and intellect (\textit{hifz al-‘aql}). This research emphasizes the need for a comprehensive Sharia audit covering contract legality, algorithmic ethics, and interface design to realize a digital economic ecosystem that is fair, transparent, and in accordance with Islamic Sharia.
\end{abstract}

\begin{IEEEkeywords} E-commerce; Fiqh Muamalah; Product; Dropshipping; Pay Later; Tadlis; Algorithmic Marketing; Halal Verification; Pre-Order; Shariah.
\end{IEEEkeywords}

\section{Introduction} \label{sec:introduction}
The development of information and communication technology (ICT) has become the epicenter of a fundamental transformation in the global economic landscape, crystallizing into a new entity known as the digital economy. Indonesia, with the world's fourth-largest population and a productive demographic bonus, is not only participating but also at the forefront of this digital revolution. This morning, Tuesday, October 30, 2025, discussions about e-commerce have undergone a fundamental shift; it is no longer merely an alternative sales channel, but has become a crucial infrastructure supporting the majority of modern society's trading activities. Projections released by Bank Indonesia (BI) consistently illustrate the exponential growth trend in e-commerce transaction value, surpassing thousands of trillions of rupiah annually \cite{BankIndonesia2022}. This figure is not merely an economic metric, but rather a manifestation of a massive and structural paradigm shift in consumer behavior from physical interactions in conventional markets to digital transactions mediated by algorithms and interfaces. This phenomenon, which is accelerating in the country with the largest Muslim population in the world, inevitably creates complex points of contact with the set of values, ethics, and laws of \textit{muamalah} contained in Islamic law.

Upon closer inspection, the rise of e-commerce is not just about having a website or shopping app. It is underpinned by a sophisticated, layered ICT architecture. The convergence of cloud computing, which provides unlimited scalability, Big Data analytics that can map consumer behavior down to the granular level, and the ubiquity of mobile internet penetration has created a hyper-efficient digital ecosystem. Within this ecosystem, technology no longer plays a passive role. Instead, artificial intelligence (AI) and machine learning (MA) are the brains behind the personalization engine, recommending products with frightening precision based on browsing history, clicks, and even the length of time a user looks at an image. Application Programming Interfaces (APIs) enable seamless integration between e-commerce platforms and financial services (FinTech), logistics, and social media, creating a seamless user experience while blurring the boundaries between transactions. This technological architecture, designed to maximize user engagement and conversion rates, inherently carries profound ethical and sharia-compliant implications.

Islam, as a comprehensive religion (\textit{syumuliyah}), is fundamentally not opposed to technological progress. The Islamic jurisprudence principle ``\textit{al-ashlu fil mu'amalah al-ibahah hatta yadulla ad-dalilu 'ala tahrimiha}'' (the basic law in muamalah is permissible until there is evidence that prohibits it) provides a very broad space for innovation. Fundamental principles such as justice (\textit{'adl}), transparency (\textit{amanah}), honesty (\textit{shiddiq}), and mutual consent (\textit{'an taradhin minkum}) \cite{Antonio2001} remain the guiding principles. However, the challenge lies in how these principles are translated in a context where contracts no longer occur directly between two humans, but are mediated by user interfaces (UI/UX) and algorithmic logic that are often ``black box'' (\textit{black box}). The pace of innovation in the e-commerce ecosystem, which produces new features almost every quarter, often outpaces ijtihad and contemporary Islamic jurisprudence. As a result, various problems, doubts (syubhat), and dilemmas arise among Muslims who strive to remain faithful to their religious teachings amidst the onslaught of digital disruption. \cite{Tarmizi2017}.

Based on in-depth observation of the technology architecture and business models in the Indonesian digital market, this study identifies at least six crucial issues that require in-depth analysis of muamalah fiqh, taking into account its ICT aspects.

First, the rise of the dropshipping business model, which is technically a form of information arbitrage enabled by digital platforms. Sellers (dropshippers) create virtual storefronts without ever owning or managing physical inventory. This practice inherently carries the risk of gharar (uncertainty) due to the information asymmetry between what is displayed in the digital storefront and the reality of stock, quality, and logistics on the supplier's side \cite{Fadila2021}. This directly challenges the principle of the prohibition of selling what one does not own (bai' ma la tamlik).

Second, the integration of FinTech products \textit{Buy Now, Pay Later} (BNPL) as a standard payment option. Technically, BNPL is an instant loan product embedded directly into the transaction flow (\textit{checkout flow}) via API. OJK data confirms the mass adoption of this service \cite{OJK2023}. The most crucial sharia-compliant issue lies in its monetization model, which is based on late fees (\textit{gharamah}). These fees, which are algorithmically calculated based on percentage and time, have characteristics identical to \textit{riba an-nasi'ah}, which is absolutely forbidden \cite{MUIJatim2022}.

Third, the gap between digital representation and the physical reality of the product (\textit{Ghabn} and \textit{Tadlis}). In the digital world, products are represented through pixels of photos, videos, and descriptive text. Manipulation of digital images and the use of ambiguous or exaggerated descriptions is very easy to do. The annual report from YLKI, which recorded a high number of complaints in this sector \cite{YLKI2024}, shows that this digital representation technology is often exploited for \textit{tadlis} (fraud) practices, which undermine the principle of buyer consent\cite{Lubis2020}.

Fourth, the use of algorithmic marketing strategies to encourage impulse purchases. Features such as flash sales with countdown timers, dynamic ``limited stock'' notifications, and AI-based recommendation engines collectively create a digital environment rife with psychological triggers. This persuasive architecture, designed to maximize conversions, risks encouraging wasteful behavior (israf), an attitude strongly condemned in Islamic teachings. \cite{Rahman2020}.

Fifth, the challenge of verifying product halal status in an open marketplace ecosystem. The platform's architecture, which allows anyone to become a seller, creates scalability challenges in terms of oversight and verification. Amidst the growing halal awareness driven by KNEKS (KNEKS2023), Muslim consumers face serious data integrity issues, as halal claims from individual sellers are difficult to independently verify, especially for critical products \cite{Zainul2022}.

Sixth, the failure of standard interfaces to facilitate Sharia-compliant pre-order (PO) contracts. PO systems are often the solution for customized products. Their fiqh equivalent, salam contracts, have strict requirements, such as full payment upfront and detailed specifications. However, standard e-commerce transaction flows generally only facilitate a down payment (DP) and a brief description, thus failing to fully fulfill the pillars and requirements of a salam contract and potentially containing gharar\cite{DSNMUI2000Salam}.

Based on a series of challenges rooted in the intersection of ICT architecture and sharia principles, this research is urgently needed. There is a lack of comprehensive guidance for Muslims in Indonesia in navigating the complexities of e-commerce. Therefore, this article has two main objectives: first, to analyze these six problematic transaction models in depth using the analytical tools of muamalah jurisprudence that considers their technological context. Second, to formulate practical guidance for Muslim consumers and sellers so that their digital economic activities are not only materially profitable but also align with sharia principles and bring blessings.

\section{Related Work} \label{sec:related-work}

Academic studies bridging the dynamics of \textit{e-commerce} and the principles of Islamic commercial jurisprudence (\textit{fiqh muamalah}) have shown significant development over the past decade. Existing literature can be classified into several thematic clusters that collectively form the foundation of this research. Early studies generally focused on fundamental questions regarding the overall validity of online transactions before shifting to more specific analyses of business models and increasingly complex transactional features.

\subsection{General Validity of E-Commerce in Fiqh Muamalah}
In the early stages of \textit{e-commerce} development, Islamic legal scholars sought to provide a juridical basis for this new form of commerce. Research by Syafruddin \cite{Syafruddin2013} is one example of an early work that analyzes \textit{e-commerce} transaction mechanisms from an Islamic legal perspective. He concludes that as long as the pillars (\textit{arkan}) and conditions (\textit{shurut}) of sale such as the presence of legally competent sellers and buyers, a clear and lawful object of transaction (\textit{ma'qud 'alaih}), and a valid offer and acceptance (\textit{sighat}) are fulfilled, the transaction is valid. In this context, offer and acceptance can be manifested through the act of clicking “buy” and confirming payment, which is considered \textit{sighat fi'liyah} (offer and acceptance through action). A similar view is reinforced by Dhinarti and Amalia \cite{Dhinarti2019}, who assert that digital platforms merely serve as a new \textit{wasilah} (medium) for an already established commercial practice, without altering the essence of the sales contract. These works successfully build a theoretical foundation showing that Islam does not reject innovative transaction media as long as the substance of the contract remains intact.

\subsection{Fiqh Analysis of Specific Business Models: Dropshipping}

As business models evolved, researchers shifted their attention to more specific practices, one of which is \textit{dropshipping}. This model sparked debate because, on the surface, it appears to violate the prohibition against selling goods that one does not yet own. Fadila and Sartika \cite{Fadila2021} highlight the potentially high \textit{gharar} (uncertainty) in \textit{dropshipping}, particularly regarding stock availability, product quality, and delivery time, since the seller (dropshipper) does not have direct control over the goods. Tarmizi and Hamzah \cite{Tarmizi2021} deepen this analysis by examining various \textit{dropshipping} schemes and conclude that the most problematic are those without a clear contract between the dropshipper and the supplier. These academic studies, which highlight risks and seek Sharia-compliant solutions, contributed to a discourse that was eventually addressed by the fatwa council. The National Sharia Council of the Indonesian Ulema Council (DSN-MUI) later issued Fatwa No. 145/DSN-MUI/XII/2021 \cite{DSNMUI2021Dropship}, which provides a solution by permitting \textit{dropshipping} if conducted through a \textit{wakalah bil ujrah} (agency with fee) or \textit{salam} (forward sale) contract. This demonstrates a healthy cycle between academic research and contemporary fatwa formulation.

\subsection{Critical Studies on Digital Financing Features: Buy Now, Pay Later (BNPL)}
The emergence of digital financing instruments such as BNPL or PayLater has triggered a new wave of highly critical research. Although these features offer convenience, they carry serious Sharia implications. Aritonang \cite{Aritonang2022} conducted an in-depth case study of the Shopee PayLater service, examining its cost and late-fee structure, and concludes that the penalty meets the criteria of usury (riba). Likewise, Wafa \cite{Wafa2020} in her review of similar services reaches the same conclusion, asserting that imposing additional charges due to late debt repayment constitutes \textit{riba an-nasi'ah}. More recent research by Putra et al. \cite{Putra2025} strengthens these findings by analyzing other PayLater products on the market. These academic arguments strongly support the stance of religious institutions, such as the East Java MUI Fatwa \cite{MUIJatim2022}, which explicitly prohibits conventional PayLater schemes that impose penalties. As a contrast, researchers often refer to DSN-MUI Fatwa No. 117/DSN-MUI/II/2018 on Sharia-Based Information Technology Financing Services \cite{DSNMUI2018Fintech} as a blueprint for designing digital financing services that avoid elements of riba.

\subsection{Consumer Protection and Ethical Concerns}
Issues of consumer protection from unfair market practices have also received substantial attention. Lubis and Ismaulina \cite{Lubis2020} specifically examine \textit{tadlis} (fraud through inaccurate descriptions) in online business, identifying various modus operandi of dishonest sellers and its impact on contract validity. Meanwhile, Nofriandi \cite{Nofriandi2021} takes a broader approach by analyzing various forms of \textit{gharar} in \textit{e-commerce} transactions, from unclear product specifications to uncertainty in the delivery process. From the consumer behavior perspective, research has begun exploring the psychological impact of digital marketing strategies. Huda and Huda \cite{Huda2020} analyze the phenomenon of \textit{flash sales} from an Islamic legal standpoint, weighing the benefits of discounts against the risk of manipulative tactics that encourage unnecessary purchases. Furthermore, Rahman \cite{Rahman2020} explicitly links the convenience and intensity of online promotion to the emergence of wasteful behavior (\textit{israf}), an ethical-moral issue that goes beyond mere contract validity.

\subsection{Product Integrity and Order-Based Contracts}
Two other crucial areas are halal product assurance and the validity of order-based contracts. Zainul et al. \cite{Zainul2022} highlight the significant challenges in building a reliable halal product assurance ecosystem within open and decentralized \textit{e-commerce} platforms. They emphasize the information gaps and verification difficulties faced by Muslim consumers. On the other hand, the practice of \textit{pre-order} (PO) is examined in detail by Mubarok and Hasan \cite{Mubarok2019}. Through case studies, they find that many PO practices fail to meet the strict requirements of \textit{salam} or \textit{istishna'} contracts as stipulated in DSN-MUI fatwas \cite{DSNMUI2000Salam, DSNMUI2000Istishna}. Failure to establish definite specifications, full upfront payment, and clear delivery timelines renders many PO transactions potentially invalid according to fiqh standards.

\subsection{Research Positioning and Contribution}
Based on the above literature mapping, it is evident that researchers have made significant contributions in analyzing various aspects of \textit{e-commerce} transactions from an Islamic perspective. However, most of these studies tend to be fragmentary, focusing on one or two issues separately for example, discussing only \textit{dropshipping} or analyzing only PayLater. To date, there remains a scarcity of studies that provide a comprehensive and integrated analysis that simultaneously addresses the six key problems most frequently faced by Muslim consumers and business actors in Indonesia. This research aims to fill that gap. By integrating analyses of \textit{dropshipping}, BNPL, \textit{tadlis}, \textit{israf}, halal assurance, and \textit{pre-order} into a single, coherent \textit{fiqh muamalah} framework, this study offers a holistic guide that not only identifies problems but also formulates practical solutions relevant to the current digital marketplace in Indonesia.

 
\section{Methodology} \label{sec:methodology}
\begin{figure}[!ht]
    \centering
    \includegraphics[width=\columnwidth]{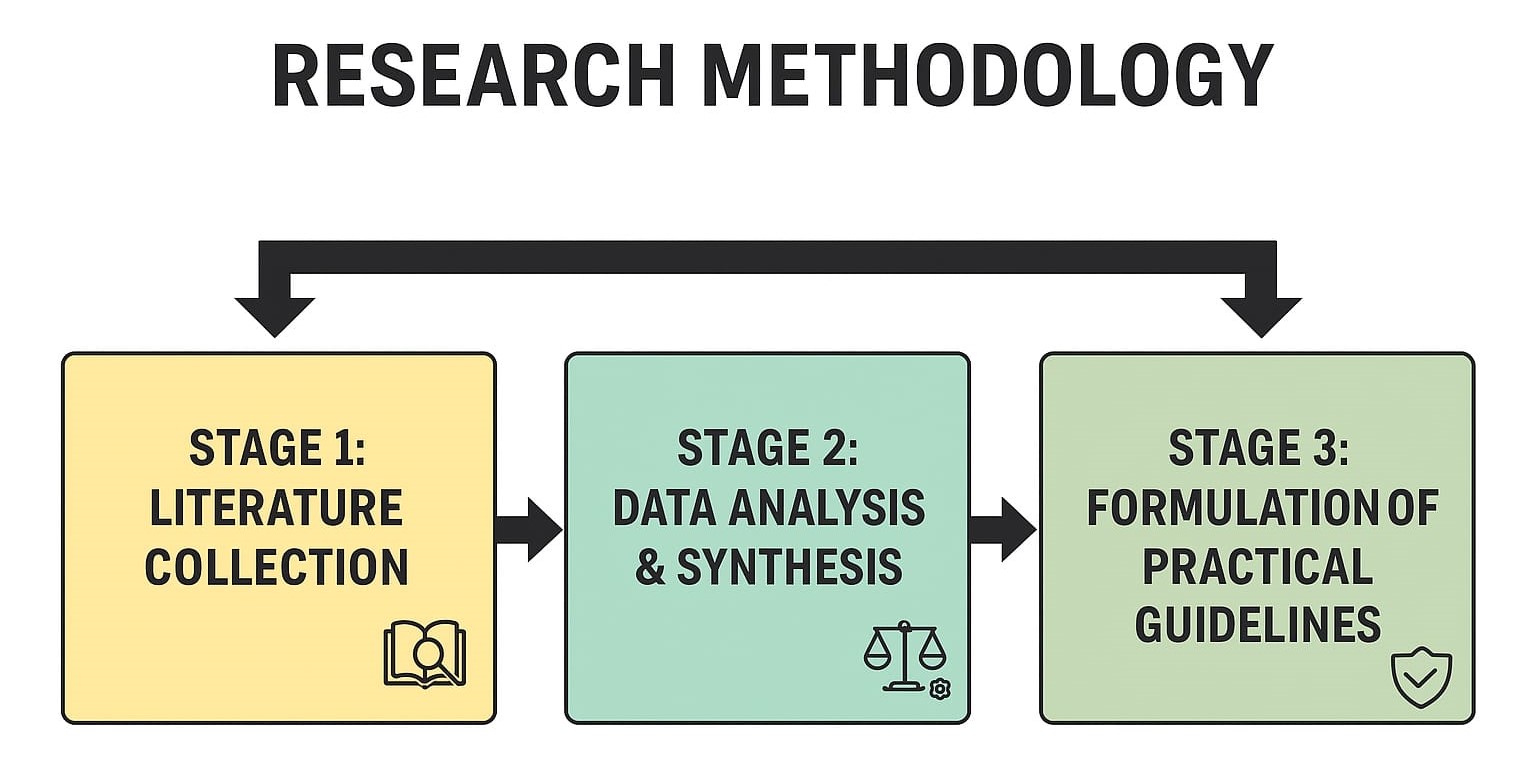}
    
    \caption{Flowchart diagram.}
    \label{fig:metodologi_flowchart} 
\end{figure}

Fundamentally, this research is grounded in a Critical Literature Review methodology utilizing a normative juridical approach. The selection of this methodology is strategic, predicated on the evaluative and normative nature of the research inquiry, which seeks to audit the compliance of contemporary e-commerce transactional practices against the norms, precepts, and principles enshrined in Islamic law (Fiqh Muamalah). This study does not attempt to measure quantitative data or empirical phenomena; rather, it aims to construct a robust and justifiable legal argumentation based on an in-depth analysis of authoritative textual sources. The research process is systematically designed into three continuous stages, as illustrated in Fig.~\ref{fig:metodologi_flowchart}.

\subsection{Literature Sourcing Stage (Investigative Data Collection)}
This phase constitutes an investigative process and the compilation of secondary data, serving as the material foundation for the entire analysis. The collected sources are not treated equally; instead, they are hierarchically classified based on their role and authority in constructing Islamic legal arguments:

\begin{enumerate}
\item {Authoritative Primary Sources (Dalil Naqli)}
This category represents the supreme legal foundation and functions as the constitution within this research framework. These sources include:
\begin{enumerate}
\item The Holy Qur'an, which provides universal principles and fundamental prohibitions (\textit{qath'iyyah al-dalalah}), such as the absolute prohibition of usury (QS. Al-Baqarah: 275), the ban on acquiring wealth through vanity, and the necessity of mutual consent (\textit{'an taradhin minkum}).
\item Al-Hadith Ash-Sharif, offering elaboration, technical details, and applicative examples of Quranic principles. Relevant Hadiths, such as the prohibition of selling unowned goods (\textit{bai' ma la tamlik}) and the ban on transactions containing excessive uncertainty (\textit{gharar}), serve as primary references.
\end{enumerate}

\item {Secondary Juridical Sources (Fiqh Interpretation and Ijtihad)}
Functioning as jurisprudence, legal doctrine, and implementing regulations that interpret primary sources. This category includes:
\begin{enumerate}
\item Classical and contemporary Fiqh Muamalah texts \cite{lZuhayli2011}, providing technical definitions, precise terminology, as well as the pillars and conditions of various contracts. These literatures provide an established analytical framework and ``legal dictionary''.
\item Fatwas from credible institutions, specifically the National Sharia Council - Indonesian Ulema Council (DSN-MUI) \cite{DSNMUI2021Dropship, DSNMUI2000Salam}. These fatwas represent collective reasoning (\textit{ijtihad jama'i}) of scholars in responding to the modern Indonesian socio-economic context, thus possessing high contextual relevance.
\item Contemporary muamalah analysis books and treatises \cite{Tarmizi2017}, representing individual reasoning (\textit{ijtihad fardi}) from Islamic economic experts. These sources often offer more agile and in-depth analyses of recent market innovations not yet covered by formal fatwas.
\end{enumerate}

\item {Contextual and Empirical Sources (Field Facts)}
Functioning as the ``official record of proceedings'' providing objective descriptions and supporting data regarding the studied phenomena. These sources ensure that the Fiqh analysis is grounded in Indonesian practice. This includes academic journals \cite{Fadila2021, Aritonang2022}, as well as factual reports from institutions such as Bank Indonesia (BI) \cite{BankIndonesia2022}, the Financial Services Authority (OJK) \cite{OJK2023}, and the Indonesian Consumers Foundation (YLKI) \cite{YLKI2024}.

\end{enumerate}

\subsection{Data Analysis and Synthesis Stage (Juridical Analysis and Synthesis)}
This is the core of the research process, where all collected sources are processed through an ``intellectual trial process'' consisting of four logical steps:

\begin{enumerate}
\item {Description of the Phenomenon}
Based on contextual sources, each transaction model under study is described in detail and objectively. Technical mechanisms, process flows, as well as the rights and obligations of the involved parties are presented clearly, free from initial judgment.
\item {Identification of the Sharia Framework}
For each phenomenon, identification, mapping, and elaboration of proofs, legal maxims (\textit{qawa'id fiqhiyyah}), and relevant muamalah principles from primary and secondary sources are conducted. This stage aims to build an ideal legal framework as a standard for evaluation.
\item {Critical Analysis (Dialectical Comparison)}
At this stage, the facts derived from the phenomenon description (step 1) are dialectically juxtaposed against the ideal Sharia framework (step 2). This process is critical in nature, interrogating and evaluating each transaction component to identify points of compliance as well as points of divergence or violation regarding Islamic legal norms.
\item {Synthesis and Legal Conclusions (\textit{Istinbath al-Hukm})}
The results of the critical analysis are subsequently synthesized to draw an argumentative legal conclusion. This \textit{Istinbath al-Hukm} process yields not only a final legal verdict (e.g., valid, invalid, or \textit{syubhat}), but more importantly, provides a detailed elucidation of the legal reasoning (\textit{'illah al-hukm}) and precisely identifies the locus of any problematic elements.
\end{enumerate}

\subsection{Applicative Guideline Formulation Stage}
The final phase of this research is fundamentally designed to transcend theoretical conclusions and generate practical impact (\textit{maslahah}). The primary focus of this stage is to transform the entire synthesis and legal deductions (\textit{Istinbath al-Hukm}) obtained from the critical analysis of the six e-commerce phenomena into an applicative guide. This transformation process involves translating complex Fiqh analysis into recommendations that can be directly implemented by stakeholders, thereby bridging the gap between Islamic legal theory and daily digital commercial practices.

The guide is specifically crafted using clear, straightforward, and simple language, avoiding intricate technical Fiqh terminology to ensure accessibility for its primary target audience: Muslim consumers and sellers. The formulation of this guide focuses on delivering concrete and actionable recommendations. Rather than merely presenting prohibitions, this guide offers specific steps users can take to verify transactions, identify Sharia risks, or modify their business models to align with Muamalah precepts.

The ultimate objective of this formulation is to empower users to navigate the complexities of the e-commerce ecosystem with greater confidence, security, and alignment with Sharia principles. This guide will not only highlight practices to be avoided but will also proactively suggest and explain alternative schemes permissible under Sharia. For instance, rather than merely prohibiting specific dropshipping schemes, the guide will elucidate how to implement \textit{wakalah} or \textit{salam} contracts as valid solutions.

\section{Result and Discussion} \label{sec:result}
This study examines six crucial issues in modern e-commerce
that lie at the intersection of disruptive technological innovation and the permanent principles of \textit{Fiqh Muamalah}. Each issue is analyzed using a systematic three-step methodological framework: an in-depth description of the technological phenomenon (\textit{Tashwir al-Mas'alah}) to map out the business processes and technical architecture, identification of the relevant Sharia framework (\textit{Takyif al-Fiqhi}) by referring to the principles of naqli and the rules of jurisprudence, critical analysis to find points of contact, potential violations, and \textit{'illat} (legal reasons) behind them, and to textual evidence and jurisprudential principles, critical analysis to find points of contact, potential violations, and the underlying \textit{'illat} (legal rationale), and concluding with the derivation of legal conclusions (\textit{Istinbath al-Hukm}) along with proposed solutions.

\subsection{Fiqh Muamalah Analysis of the Dropshipping Business Model}

\begin{enumerate}
    \item Description of the Phenomenon

    The \textit{dropshipping} business model is technically a form of supply chain virtualization and information arbitrage facilitated by digital platform architecture \cite{Fadila2021}. In practice, the seller (\textit{dropshipper}) creates a virtual storefront to market a product without ever owning or managing the physical inventory of the goods \cite{Fadila2021, DSNMUI2021Dropship}. The \textit{Dropshipper} operates as a \textit{capital-light} entity, wherein inventory, warehousing, and logistics risks are fully assumed by a third party (the supplier). The transaction flow begins with the consumer placing an order and making a payment to the \textit{dropshipper} through their digital storefront. Then, the \textit{Dropshipper}, after taking a profit margin, forwards the order details and payment (at the cost price) to the supplier. Subsequently, the supplier packs and ships the goods directly to the consumer's address, often on behalf of the \textit{dropshipper}. In this scheme, the \textit{dropshipper} never holds physical \textit{qabd} (possession) nor \textit{hukmi} (constructive/legal) possession over the goods at any point in the transaction flow.
    
    \item Identification of the Sharia Framework
    
    This practice directly conflicts with one of the fundamental principles in Fiqh Muamalah, namely the prohibition of selling something that is not yet owned or possessed (\textit{bai' ma la tamlik}). This principle stems from very clear Hadith evidence, narrated from Hakim bin Hizam who asked the Messenger of Allah (PBUH), O Messenger of Allah, someone comes to me wanting to buy goods that I do not yet possess. May I sell them, then buy them from the marketplace?`` The Messenger of Allah (PBUH) replied, Do not sell what you do not possess (\textit{Laa tabi' ma laysa 'indaka})''. (HR. Tirmidzi, Abu Daud). The \textit{'Illat} (legal rationale) behind this prohibition is to prevent excessive \textit{gharar} (uncertainty) and \textit{maysir} (speculation) \cite{Fadila2021, Tarmizi2017}. \textit{Gharar} in \textit{dropshipping} becomes highly relevant due to the high information asymmetry between what is displayed on the digital storefront and the reality of stock, quality, and logistics on the supplier's side, which are entirely outside the \textit{dropshipper}'s control. Furthermore, this violates a fundamental legal maxim: \textit{Al-kharaj bi al-daman} (Revenue/profit accompanies risk/liability). The \textit{Dropshipper} takes full profit (\textit{kharaj}) from the sale without ever bearing the risk of ownership (\textit{daman}) over the goods.

    \item Critical Analysis

    Based on \textit{Takyif al-Fiqhi} (fiqh adaptation), if the contract used by the \textit{dropshipper} is a pure sale contract (\textit{bai'}), then the transaction has a high potential to become fasad (corrupt/invalid) \cite{Tarmizi2021}. This is because the \textit{dropshipper} (as the seller) does not fulfill the condition of \textit{in'iqad} (contract formation), namely ownership (\textit{milkiyyah}) or at least possession (\textit{qabd}) of the \textit{ma'qud 'alaih} (object of the contract) when the contract takes place. Without \textit{qabd}, the \textit{dhaman} (liability for risk) for the goods does not transfer to them, whereas the transfer of risk is the essence of buying and selling. Existing scientific literature confirms that this high potential for \textit{gharar} constitutes the core of the Sharia issue \cite{Fadila2021}. Uncertainty regarding stock availability (whether the goods exist or not), quality assurance (whether it matches the specifications in the storefront), and timeliness of delivery because the \textit{dropshipper} has no direct control over the goods constitute a violation of the principles of transparency and fairness in transactions. Previous research also highlights that this problem becomes increasingly complex because it is often not based on a clear contract between the \textit{dropshipper} and the supplier themselves, causing the status of the \textit{dropshipper} to become ambiguous: they are neither a seller (because they do not own the goods) nor an official agent (because there is no \textit{wakalah} contract) \cite{Tarmizi2021}.

    \item Synthesis and Legal Conclusions

    Responding to academic discourse and practical needs in society, the National Sharia Board of the Indonesian Council of Ulama (DSN-MUI), through Fatwa No. 145/DSN-MUI/XII/2021, provides a solution (\textit{makhraj}) to this legal impasse \cite{DSNMUI2021Dropship}. \textit{Dropshipping} can be justified according to Sharia (valid) if its contract scheme is reformulated clearly and transparently. The \textit{Istinbath al-Hukm} (legal conclusion) drawn is:

    \begin{enumerate}

        \item The practice of \textit{dropshipping} using a direct sale contract (\textit{bai'}), where the \textit{dropshipper} acts as the seller of goods they do not own, is invalid. This is because it violates the condition of ownership (\textit{bai' ma la tamlik}) and contains high \textit{gharar} \cite{DSNMUI2021Dropship}.
        \item This practice becomes valid if the contract is changed to an agency contract (\textit{wakalah bil ujrah}). In this scheme, the \textit{dropshipper} must clearly position themselves as a representative (intermediary/agent) of the supplier to market the goods, or as a representative of the buyer to source the goods. For their services, the \textit{dropshipper} is entitled to receive a clear and agreed-upon wage (\textit{ujrah} or \textit{fee}), rather than taking a ``profit margin'' from a sale \cite{DSNMUI2021Dropship}.
        \item As another alternative, \textit{dropshipping} can be valid if using the \textit{salam} contract scheme (sale by order). In this scheme, the \textit{dropshipper} acts as the seller (\textit{muslam ilaih}) who accepts an order with specifications (\textit{al-wasf al-mundhabit}) and a price that are clear at the beginning, and the capital (payment) is received in full in advance (\textit{ta'jil ra's al-mal}) from the consumer (\textit{rabb al-salam}). The \textit{Dropshipper} then uses the money to buy the goods from the supplier and requests them to ship to the consumer \cite{DSNMUI2021Dropship, DSNMUI2000Salam}.
    
    \end{enumerate}
\end{enumerate}

\subsection{Analysis of Islamic Jurisprudence on the Buy Now, Pay Later (BNPL) Feature}

\begin{enumerate}
    \item Description of the Phenomenon
    
    The Buy Now, Pay Later (BNPL) feature is an instant credit product embedded directly into the checkout flow through API integration \cite{OJK2023}. It is a form of embedded finance that eliminates the friction of traditional credit applications. Within the framework of Islamic jurisprudence (Fiqh Muamalah), BNPL can be implemented through two main schemes: the Qardh (Loan) Scheme and the Bay' Muajjal (Deferred Sale and Purchase) Scheme. The Qardh (Loan) Scheme is where BNPL providers (P2P Lending or multifinance) provide loans (cash advances) to consumers to pay for goods to merchants \cite{Aritonang2022}. Consumers then repay the debt to the BNPL provider in installments. Then there is the \textit{Bay' Muajjal} (Tough Buying and Selling) scheme, where the BNPL provider ``buys'' goods from the \textit{merchant} (often legally) and then ``sells'' them back to the consumer at a higher price (including margin) paid in installments.

    \item Identification of the Sharia Framework

    The problem with Islamic law arises not from the concept of \textit{Bay’ Muajjal} (which is essentially permissible), but from the additional components that accompany it. The central issues are \textit{Riba} and \textit{Gharar}.
    \begin{enumerate}
        \item Usury, which is a violation that manifests itself in two forms. First, \textit{Riba Qardh}, which is any additional benefit required on top of the principal loan. This is based on the agreed-upon fiqh principle: ``\textit{Kullu qardin jarra manfa'atan fahuwa riba}'' (Every loan that brings benefit [to the lender] is riba). This manifests itself in ``service fees'' or ``administrative fees'' calculated as a percentage of the principal loan, not as \textit{ujrah} on a fixed real cost (\textit{fixed fee}) \cite{Wafa2020}. Second, Riba Nasiah (riba due to delay), which most commonly and clearly occurs through Late Payment Fees (\textit{Late Payment Fee}) \cite{MUIJatim2022}. If this fine is calculated based on a percentage of the remaining debt and accumulated and becomes the income of the BNPL provider, it is identical to the practice of usury of the Jahiliyah era which is definitively forbidden by the Qur'an (Q.S. Al-Baqarah: 275-279) \cite{Aritonang2022}.
        
        \item Gharar, which is something that arises from hidden terms and conditions (\textit{hidden fees}), or mixing contracts (\textit{akadun fi akadain}) between buying and selling and loans without a clear separation, which is prohibited in the Hadith.
    \end{enumerate}
    
    DSN-MUI Fatwa No. 117/DSN-MUI/II/2018 concerning Sharia Information Technology-Based Financing serves as a reference that financing services must be free from usury elements\cite{DSNMUI2018Fintech}.

    \item Critical Analysis

    The fundamental difference between Conventional and Sharia BNPL lies in the source of profit. Conventional BNPL is rooted in interest-bearing loans (Qardh), where profits are derived from remuneration for the time and amount of money lent (Riba) \cite{Putra2025}. This is evident in the late fees that are accumulative and become company income, which is clearly Riba Nasiah \cite{MUIJatim2022}. In contrast, Sharia BNPL must use a Deferred Sale and Purchase (Bay' Muajjal or Murabahah) contract, where the profit (Ribh) is the difference between the purchase price and the selling price agreed upon in advance and is fixed, regardless of when payment is made (as long as it is within the deadline). Late payment fines in the sharia scheme (if any) must be in the form of \textit{Ta'zir} (disciplinary sanctions) which must not be accumulative and must be distributed as social/charity funds (\textit{al-qardh al-hasan funds}), not as company income \cite{DSNMUI2018Fintech}. The \textit{ujrah} scheme for services (if using the \textit{qardh} contract) must also be in the form of a fixed real fee, not a percentage.

    \item Synthesis and Legal Conclusions

    The basic concept of deferred sale (Bay' Muajjal), where the deferred price is higher than the cash price, is valid and permissible (the principle of al-tsaman li al-ajal), according to the majority of scholars \cite{Antonio2001}.

    The majority of scholars and fatwa institutions in Indonesia, including the LBM PBNU and the East Java MUI \cite{MUIJatim2022}, prohibit conventional BNPL due to the late fees that are riba-like and/or service fees calculated based on a percentage of the principal debt (riba qardh).
    
    BNPL will only comply with Islamic jurisprudence if it is operated purely as a sharia-compliant Bay' Muajjal (murabahah li al-amir bi al-syira contract) with a fixed margin and non-profitable ta'zir fees. Alternatively, if using the \textit{qardh} contract, the service fee (\textit{ujrah}) must be fixed (\textit{fixed cost}) to cover real operational costs, and there must be no fines that become income \cite{DSNMUI2018Fintech}.
    
\end{enumerate}

\subsection{Fiqh Muamalah Analysis of Tadlis and Ghabn (Digital Representation Discrepancy)}

\begin{enumerate}
    \item {Description of the Phenomenon}
    
    In the digital realm, products are represented exclusively through pixels, video footage, and textual descriptions \cite{DSNMUI2021Dropship}. Unlike physical marketplaces where \textit{mu'ayyanah} (physical inspection) and \textit{ru'yah} (direct viewing) are feasible, the digital market is contingent upon absolute reliance on visual and textual representations. Technologies such as image editing software, color grading, and persuasive copywriting enable sellers to manipulate digital imagery and utilize ambiguous or exaggerated descriptions (overclaiming). This creates a significant gap or disparity between the expectations constructed in the consumer's mind and the physical reality of the product received. The high volume of consumer complaints in this sector, as recorded by YLKI \cite{Tarmizi2017}, indicates that these technologies are frequently exploited for detrimental practices.

    \item {Identification of the Sharia Framework}
    
    This phenomenon directly transgresses two fundamental prohibitions in \textit{muamalah}. First is \textit{tadlis} (fraud or deceptive information), specifically \textit{kitman al-'aib} (concealing defects) or \textit{izhhar al-mahasin} (feigning non-existent virtues) regarding the object of the contract to mislead the buyer \cite{Putra2024}. This is strictly forbidden by the Prophet SAW, who stated: ``Whoever deceives us is not one of us'' (HR. Muslim).

    Second is \textit{ghabn} (unjust loss), which refers to an extreme imbalance between the price paid (\textit{tsaman}) and the actual value of the goods (\textit{mutsman}). Islamic jurisprudence distinguishes between \textit{ghabn yasir} (minor, tolerable loss) and \textit{ghabn f\={a}\d{h}ish} (excessive loss beyond customary limits). When \textit{ghabn f\={a}\d{h}ish} is caused by \textit{tadlis} on the part of the seller, the contract becomes \textit{fasad} (corrupt/voidable).

    These practices undermine the principle of mutual consent (\textit{'an taradhin minkum}), which is a prerequisite for the validity of a sale, in accordance with the word of Allah SWT in Q.S. An-Nisa: 29: ``O you who have believed, do not consume one another's wealth unjustly but only [in lawful] business by mutual consent'' \cite{OJK2023}.

    \item {Critical Analysis}
    
    The discrepancy in digital representation constitutes a systematic form of modern \textit{tadlis}. The consent (\textit{ridha}) granted by the buyer is legally defective (\textit{fasad al-ridha}) because it is premised on falsified information. Sellers who utilize stock photos from the internet that do not correspond to the actual item (e.g., using premium imagery for counterfeit/KW goods) or who exaggerate material specifications have intentionally created information asymmetry to secure unjust profits. E-commerce platforms facilitate this by employing features that prioritize ``visuals'' over ``substance''. According to \textit{fiqh}, a buyer subjected to \textit{tadlis} or \textit{ghabn f\={a}\d{h}ish} retains the right of \textit{khiy\={a}r} (option) to rescind the contract. In this context, the applicable mechanisms are \textit{khiy\={a}r 'aib} (option due to hidden defects), \textit{khiy\={a}r tadlis} (option due to fraud), or potentially \textit{khiy\={a}r ru'yah} (option upon inspection—specifically regarding the mismatch between description and sight), as recognized in the Hanafiyah school of thought.

    \item {Synthesis and Legal Conclusions}
    
    Transactions containing elements of \textit{tadlis} (deceptive digital representation) or resulting in \textit{ghabn f\={a}\d{h}ish} (extreme loss) for the buyer are \textit{haram} (prohibited) and invalid. The contract is deemed \textit{fasad} (corrupt) due to the violation of the principle of mutual consent (\textit{tar\={a}\d{d}in}). Consumers possess the full right of \textit{khiy\={a}r} to execute \textit{faskh} (cancellation of the transaction) and demand a full refund. E-commerce platforms bear a \textit{mas'uliyyah} (Sharia responsibility/liability) to penalize sellers proven to engage in systematic \textit{tadlis} and are obligated to facilitate the \textit{khiy\={a}r} process as a form of consumer protection.
\end{enumerate}

\subsection{Islamic Ethical Analysis of Algorithmic Marketing (Impulsive Purchasing)}

\begin{enumerate}
    \item Phenomenon Description

    Modern algorithmic marketing goes beyond merely targeted advertising; it employs \textit{Big Data} and \textit{Artificial Intelligence} (AI) to construct in-depth psychographic profiles of consumers \cite{Putra2024}. This is an architecture of persuasion manifested through \textit{``Dark Patterns''} of interface design (UI/UX) that are deliberately engineered to exploit human cognitive biases (such as \textit{loss aversion}, \textit{bandwagon effect}, or \textit{sunk cost fallacy}) and drive unwanted or irrational actions \cite{Zahratunnisa2025, Sharmila2025}. These mechanisms include \textit{Urgency Cues} (e.g., \textit{flash sales} with aggressive countdown timers) and \textit{Scarcity Cues} (e.g., ``limited stock, only 2 left'', when in fact stock remains abundant) \cite{Huda2020}. Research in behavioral economics \cite{Fadilah2025, Khairunas2020} confirms that these predictive AI recommendation algorithms and artificial time pressures significantly increase consumer vulnerability and drive impulsive purchasing, which is no longer based on rational needs (\textit{hajat}), but rather on emotional impulses (\textit{syahwat}).

    \item Identification of Shariah Framework

    The relevant shariah framework focuses on consumption ethics, prohibition of manipulation, and protection of the noble objectives of shariah (\textit{Maqashid al-Syariah}):
    
    \begin{enumerate}
    \item Prohibition of Israf and Tabzir
    
    Shariah explicitly prohibits wasteful behavior (\textit{israf}) spending on the forbidden or excessively on the permissible and squandering wealth (\textit{tabzir}). ``O children of Adam! ...Eat and drink, but do not exceed (\textit{la tusrifu}). Indeed, Allah does not like those who exceed.'' \textit{(Q.S. Al-A'raf: 31)}. And ``Do not squander (your wealth) wastefully. Indeed, the squanderers are brothers of Satan.'' \textit{(Q.S. Al-Isra: 26-27)} \cite{Mhtm2021}.
    
    \item Command of Iqtishoduna (Moderation)
    
    Islam commands rational, moderate consumption (\textit{wasathiyah}), based on \textit{maslahah} (welfare), not merely following desires (\textit{syahwat}) \cite{Kasanah2022}.
    
    \item Prohibition of Najsy and Tadlis
    
    \textit{Dark patterns} can be mapped to classical prohibitions. \textit{Fake social proof} (``50 people just bought this'') is a form of \textit{Najsy} (artificial demand manipulation to increase interest, which is forbidden by Hadith). \textit{Drip pricing} (hidden fees) is a form of \textit{Gharar} and \textit{Tadlis} \cite{DSNMUI2021, Zahratunnisa2025}.
    
    \item Maqashid al-Syariah and the Principle of Sadd al-Dzari'ah
    
    This issue touches upon two of the five primary \textit{Maqashid}: \textit{Hifdz al-Mal} (Protection of Wealth) and \textit{Hifdz al-'Aql} (Protection of Reason). The principle of \textit{Sadd al-Dzari'ah} (blocking the means to harm) becomes relevant to prohibit techniques that, although the contract may be valid, systematically encourage forbidden behavior (\textit{israf}).
    \end{enumerate}

    \item Critical Analysis

    Critical analysis reveals that the deepest ethical problem is not ``encouraging spending'', but rather ``encouraging the de-rationalization'' of consumers. The principle of \textit{Iqtishoduna} and the status of humans as \textit{mukallaf} (subjects of law) demand active participation of reason ('aql) to weigh needs (\textit{hajat}) versus desires (\textit{syahwat}) \cite{Kasanah2022}. Humans are responsible for their actions because they possess reason and \textit{ikhtiyar} (free choice). However, digital persuasion architecture, as confirmed by research \cite{Fadilah2025}, is designed to ``reduce cognitive burden'' a euphemism for ``deactivating critical faculties of reason''. This system is designed so that consumers do not need to think and only act based on emotional stimuli engineered by the platform.

    This is a direct sabotage of \textit{Maqashid Syariah}. It violates \textit{Hifdz al-Mal} (Protection of Wealth) by encouraging \textit{israf} \cite{Mhtm2021}, and more seriously, violates \textit{Hifdz al-'Aql} (Protection of Reason) by training consumers to make impulsive and reactive decisions, eliminating \textit{rusyd} (prudence) in transactions. Therefore, shariah responsibility (\textit{mas'uliyyah}) cannot be placed solely on consumers (``it's their own fault for not being able to restrain themselves''). Platforms that consciously design this exploitative ``persuasion architecture'' have systematically caused \textit{dharar} (harm) and violated the principle of \textit{``Laa dharara wa laa dhiraara''} \cite{Sharmila2025}.

    \item Synthesis and Derivation of Legal Ruling

    Based on the Maqashid and Fiqh analysis above:
    \begin{enumerate}
    \item The basic ruling on personalization that helps consumers find real needs is Mubah (Permissible) and can even have \textit{maslahah} value.
    
    \item Implementation of \textit{Dark Patterns} containing elements of \textit{Tadlis} (fake discounts, hidden fees) or \textit{Najsy} (fake demand manipulation, \textit{fake scarcity}) is Haram (Forbidden) \cite{DSNMUI2021, Sharmila2025}.
    
    \item System design (UI/UX and Algorithms) that deliberately and systematically exploits cognitive biases to drive \textit{Israf} on a massive scale is Haram (Forbidden). This is based on violations of \textit{Maqashid Syariah} (Hifdz al-Mal and Hifdz al-'Aql) and application of the principle of \textit{Sadd al-Dzari'ah}.
    
    \item The use of honest \textit{urgency} and \textit{scarcity cues} features (real discounts, real stock) has the ruling of Syubhat (Doubtful) leaning toward Makruh (Disliked). Although the contract is valid, this technique still promotes a culture of \mbox{con\-su\-me\-rism} that contradicts the spirit of \mbox{Iq\-ti\-sho\-duna}, \textit{zuhd}, and \mbox{qa\-na'ah} \cite{Kasanah2022}.
    \end{enumerate}
\end{enumerate}

\subsection{Fiqh Muamalah Analysis on Product Halal Verification}

\begin{enumerate}
    \item Description of the Phenomenon

    The modern digital \textit{marketplace} ecosystem is built on the principle of openness (\textit{open participation}) where every individual can become a seller without a rigorous curation and verification process for the products marketed. In this context, product halal claims are often conveyed unilaterally by sellers through product descriptions, without being supported by official certificate evidence from the Halal Product Assurance Organizing Agency (BPJPH) or the Indonesian Council of Ulama (MUI). This phenomenon creates information asymmetry between sellers and buyers, where Muslim consumers cannot ascertain whether the product is truly halal according to Sharia provisions and national regulations \cite{Zainul2022}. Based on the 2023 report by the National Committee for Islamic Economy and Finance (KNEKS), only a small portion of online business actors have validly integrated halal certificate numbers on \textit{e-commerce} platforms \cite{KNEKS2023}. Furthermore, the absence of system integration between the BPJPH SIHALAL \textit{database} and digital platforms causes weak credibility of halal information, opens opportunities for label manipulation, and potentially lowers public trust in the digital halal industry.

    \item Identification of the Sharia Framework

    In the view of fiqh muamalah, the clarity of the halal status of a good is an important condition for the validity of the sale and purchase contract (\textit{sihhat al-'aqd}). Unclarity or lack of assurance of halal status causes the emergence of \textit{gharar} (uncertainty) and \textit{tadlis} (misinformation/deception) regarding the object of the contract (\textit{ma'qud 'alayh}). This contradicts the basic principles of justice and transparency emphasized by Sharia. The Messenger of Allah (PBUH) said: ``Whoever deceives us is not one of us.'' (HR. Muslim, no. 101). This principle is in line with DSN-MUI Fatwa No. 146/DSN-MUI/XII/2021 which asserts that business actors are obliged to provide honest, correct, and not misleading information in every transaction \cite{DSNMUI:2022:OnlineShop}. From a maqasid perspective, the practice of halal claims without certification basis threatens two main objectives of Sharia, namely \textit{hifz al-din} (protection of religion) because it concerns compliance in consumption, and \textit{hifz al-mal} (protection of wealth) because consumers can be financially harmed due to invalid information. Therefore, the principle of precaution (\textit{ihtiyat}) and verification becomes a moral obligation for every Muslim business actor to ensure that the transactions conducted do not contain elements of \textit{syubhat} (doubt/ambiguity).

    \item Critical Analysis

    The absence of an integrated halal verification system on digital platforms indicates the existence of \textit{taqsir} (negligence) on the part of platform operators in carrying out their social and religious responsibilities, unlike conventional products such as food and beverages which are easier to check for halal status and certification, followed by certification education. Halal certification, especially for conventional products like food and beverages, plays an important role in ensuring product compliance with Islamic Sharia in this context \cite{wulandari2024proses}. \textit{Dropshippers}, resellers, and \textit{marketplace} actors hold strategic positions demanding high trust and honesty because they are the main intermediaries between producers and consumers. If a seller deliberately includes a halal claim without valid proof, they have committed severe \textit{tadlis} and taken profit from wealth obtained in a \textit{batil} (invalid/unlawful) manner. Digital platforms that continue to earn commissions from transactions of products with unclear status also potentially engage in \textit{ta'awun 'ala al-itsmi wa al-'udwan} (cooperating in sin and transgression). In the view of \textit{maqasid al-syari'ah}, this practice not only impacts individual loss but also damages the ethical order of Islamic economics and weakens public trust in the halal trade system. Therefore, the responsibility (\textit{mas'uliyyah}) of verification cannot be solely burdened on the consumer, but rather becomes a collective obligation (\textit{fardhu kifayah}) among regulators, platform operators, and business actors. API integration between the BPJPH SIHALAL system and \textit{marketplaces} is a crucial step to close the gap of \textit{gharar} and restore justice (\textit{adl}) in online transactions \cite{Zainul2022}.

    \item Synthesis and Legal Conclusions

    Based on the analysis above, the practice of buying and selling with halal claims that cannot be validly verified contains elements of \textit{tadlis} (fraud/deception) and \textit{gharar} (uncertainty) regarding the object of the contract, thus the ruling is \textit{haram} and the contract is classified as \textit{fasad} (corrupt/invalid). Products with unclear halal status fall into the category of \textit{syubhat}, which must be avoided by every Muslim. \textit{E-commerce} platforms in Muslim-majority countries have a Sharia responsibility (\textit{mas'uliyyah}) to ensure product data connectivity with the national halal certification system. This effort is not merely the fulfillment of administrative regulations, but a form of implementation of the values of \textit{sidq} (honesty), \textit{amanah} (trust/responsibility), and \textit{maslahah} (public benefit) which are the core of maqasid sharia. Thus, the digitalization of the halal verification system is a strategic step in maintaining the sustainability of a Sharia economic ecosystem that is fair, transparent, and oriented towards blessings. \end{enumerate}

\subsection{Fiqh Muamalah Analysis of Pre-Order (PO) Contracts}

\begin{enumerate}
    \item Description of the Phenomenon
    
    The pre-order (PO) system is a business solution for customized products, imported products with complex logistics, or products that are made to order \cite{Mubarok2019}. In this scheme, the contract is made before the goods exist or are controlled by the seller. The problem is that standard e-commerce transaction flows (UI/UX) often only facilitate down payments (DP) and brief descriptions. This flow is designed for spot sales, not for order contracts that have strict conditions and requirements in Islamic jurisprudence. As a result, many PO practices fail to meet the relevant Sharia contract criteria.

    \item Identification of the Sharia Framework
    
    PO is a form of transaction in which goods are delivered at a later date (\textit{mu'ajjal}) while the goods do not yet exist (\textit{ma'dum}). To avoid the prohibition of \textit{bai' ma la tamlik} and \textit{bai' al-ma'dum} (selling what does not exist), a PO must be classified (\textit{takyif}) into one of the following two order contracts:
    \begin{enumerate}
        \item \textit{Bai' Salam}, which is a contract for goods with specific specifications (\textit{al-wasf al-mundhabit}) with the condition of full payment in advance at the time of the contract (\textit{ta'jil ra's al-mal}) and a clear delivery time (\textit{ajal ma'lum}) (Fatwa DSN-MUI No. 05/DSN-MUI/IV/2000) \cite{DSNMUI2000Salam}.
    
        \item \textit{Bai' Istishna'}, which is a contract for the manufacture of goods if the goods need to be made/produced (e.g., custom furniture, tailored clothing). In istishna', the specifications must be clear, but the payment is flexible; it can be made in advance, in installments, or at the end, as agreed (DSN-MUI Fatwa No. 06/DSN-MUI/IV/2000) \cite{DSNMUI2000Istishna}.
    \end{enumerate}

    Failure to fulfill these conditions, especially the condition of full payment in advance for salam, makes the contract contain gharar \cite{Mubarok2019} and potentially fall into other prohibitions.

    \item Critical Analysis
    
    Many PO practices in e-commerce fail to fulfill the conditions of salam and istishna'. Critical analysis finds two main problems; First, if the platform facilitates a down payment (DP) for goods that are *not* in the \textit{istishna'} category (for example, PO for imported shoes that are finished goods, not made to order), then the “full payment in advance” requirement of the \textit{salam} contract is not met. Then second, when the \textit{salam} requirement is not met (payment is deferred/DP), this transaction falls into the prohibited category of \textit{bai' al-kali' bi al-kali'} (selling debt with debt). The buyer owes the remaining payment, and the seller owes the delivery of the goods. This is prohibited by ijma' (consensus) of the scholars. In addition, the lack of detailed product descriptions and uncertain delivery times (e.g., “estimated 4-8 weeks”) add an element of \textit{gharar} which is prohibited. Therefore, standard \textit{e-commerce} interfaces are often insufficient in terms of fiqh to facilitate sharia-compliant PO contracts.

    \item Synthesis and Legal Conclusions
    
    The practice of \textit{Pre-Order} (PO) is valid if and only if it fulfills the conditions and requirements of one of two contracts, namely \textit{Bai' Salam} (full payment in advance, clear specifications, clear delivery time) for finished goods/commodities, or \textit{Bai' Istishna'} (contract for the manufacture of goods with clear specifications, flexible payment) for goods that need to be produced. If the PO practice on the platform only uses a down payment (not full payment) for finished goods (non-Istishna'), then the contract is invalid. This is because it violates the \textit{ta'jil ra's al-mal} requirement of the \textit{salam} contract and falls under the prohibition of \textit{bai' al-kali' bi al-kali'}, as well as containing a high degree of \textit{gharar}.

\end{enumerate}

\subsection{Formulation of Practical Guidelines for Muslim Consumers and Sellers}

In accordance with the final stage of the research methodology, this section transforms the legal conclusions (\textit{istinbath al-hukm}) that have been analyzed in sub-chapters A to F into a practical guide. This guide is designed using clear and applicable language to assist Muslim consumers and sellers in navigating the \textit{e-commerce} ecosystem to align with Sharia principles.

\begin{enumerate}
    \item Guide for Muslim Consumers

    For consumers, caution (\textit{ihtiyath}) and sharia digital literacy are key.

    \begin{enumerate}
        \item Regarding \textit{Buy Now, Pay Later}, consumers are obliged to avoid using conventional BNPL or \textit{Pay Later} services that apply late fees on debt. Based on the analysis (Sub-Chapter B), these fines possess characteristics identical to \textit{riba an-nasi'ah}. As a permissible alternative, consumers can seek digital financing services that are Sharia-based and registered with the OJK, which use non-usurious contracts in accordance with DSN-MUI Fatwa.
        
        \item Regarding Product Quality (\textit{Tadlis dan Ghabn}), consumers must be critical and not easily deceived by excessive digital representations (photos/videos). Always verify by reading reviews from other buyers, checking the seller's reputation, and understanding the \textit{return policy} as a form of protection for \textit{khiyar} (option) rights in the event of a discrepancy.
        
        \item Regarding \textit{Pre-Order} (PO), when engaging in this transaction, consumers must ensure that the transaction fulfills the pillars of the \textit{Salam} contract. Ensure that product specifications (material, size, quality) are explained in great detail and in writing. Understand that for a \textit{Salam} contract to be valid, ideally, payment is made in full in advance. Avoid POs where the description is ambiguous and the payment scheme is unclear as they carry a high risk of containing \textit{gharar}.
        
        \item Regarding halal verification, consumers have an active obligation to verify products such as food, beverages, medicines, and cosmetics. Prioritize purchasing from sellers with \textit{Official Store} status or those who clearly display a valid Halal logo and certification number from BPJPH.
        
        \item Regarding excessive consumption behavior (\textit{Israf}), consumers must guard themselves against the platform's ``persuasion architecture'' that encourages impulsive buying, such as \textit{flash sale}s or stock notifications. Apply Islamic consumption principles by distinguishing between needs (\textit{hajat}) and desires (\textit{syahwat}) to avoid wasteful behavior (\textit{israf}) which is condemned.
        
    \end{enumerate}

    \item Guide for Muslim Sellers

    For sellers, the principles of honesty (\textit{shiddiq}) and transparency (\textit{amanah}) are the foundation for building a blessed business.

    \begin{enumerate}
    \item Regarding \textit{Dropshipping}

    Sellers are prohibited from using a pure sale contract for goods they do not own. For the transaction to be valid, the seller must clearly reformulate the contract. Use one of the two alternative schemes allowed by the DSN-MUI Fatwa, namely the wakalah bil ujrah contract and the salam contract. The \textit{Wakalah bil Ujrah} Contract`` means the seller acts as a representative (agent/broker) of the supplier and is entitled to a wage (\textit{fee}) for marketing services. Meanwhile, The \textit{Salam} Contract'' is where the seller acts as a full seller based on an order, where they receive full payment in advance from the buyer, then they buy the goods from the supplier to fulfill the order.

    \item Regarding Product Description \textit{(Tadlis)}

    Sellers are obliged to provide an honest and accurate product description. Use original product photos or videos (real-pict) and explain the specifications as they are. Hiding product defects or exaggerating quality is a form of \textit{tadlis} which is haram and undermines the validity of the contract.

    \item Regarding \textit{Pre-Order}

    Sellers must ensure the offered PO contract complies with Sharia. If using the \textit{Salam} contract, ensure payment is received in full in advance and the specifications as well as the delivery time are very clear. If the goods are customized (custom-made), the seller can use the \textit{Istishna'} contract, which allows for staggered payments (including DP), but product specifications must still be detailed and agreed upon at the beginning.

    \item Regarding Halal Product Guarantee

    Sellers dealing in critical products have a moral and Sharia responsibility to guarantee the halal status of their products. Displaying an official Halal certificate is a form of transparency and honesty that is absolutely necessary to protect Muslim consumers from doubt (\textit{syubhat}). 
    
    \end{enumerate}
\end{enumerate}

\section{Conclusion} \label{sec:conclusion}

This research asserts that the development of transaction systems on \textit{e-commerce} platforms not only brings convenience to the buying and selling activities of the Muslim community but also gives rise to new challenges requiring serious attention from the perspective of Fiqh Muamalah. In general, digital technology and online platforms are means that are \textit{mubāh} (permissible); however, the features and business models developing within them have the potential to contain practices that are not aligned with Sharia principles if not examined within the appropriate legal framework.

From the analysis conducted, there are three main conclusion points. First, transaction models based on information arbitrage such as \textit{dropshipping} and \textit{pre-order (PO)} cannot be declared valid if using a direct sale contract (\textit{bai'}), due to the non-fulfillment of the element of possession (\textit{qabd}) and the presence of high potential for \textit{gharar}. In PO practices with a down payment (DP) system for goods that do not yet exist (\textit{ma'dum}), the transaction risks falling under the prohibition of \textit{bai' al-kali' bi al-kali'}. Nevertheless, both models can be accommodated according to Sharia if reformulated using the appropriate contracts, such as \textit{wakālah bil ujrah}, \textit{salam}, or \textit{istishna'} in accordance with their pillars and conditions.

Second, conventional BNPL (Buy Now, Pay Later) financing schemes are categorized as non-compliant with Sharia principles, primarily due to the presence of late fees which act as additions to the principal debt and include the characteristics of \textit{Riba Nasiah}, as well as percentage-based service fees which fall into the category of \textit{Riba Qardh}. Thus, BNPL schemes can only be justified if they use contracts that are truly free from elements of riba and apply penalty mechanisms that are social, not commercial, in nature.

Third, the practice of \textit{tadlis} in the form of disguising quality or inappropriate digital product representation, including halal claims without valid verification, is considered to damage the element of mutual consent (\textit{tarādhī}) and contradicts the protection of Muslim consumer rights, particularly in the aspects of \textit{hifz al-din} and \textit{hifz al-mal}. This underscores the importance of transparency, honesty, and halal verification procedures that are legally and ethically accountable.

The most important finding of this research indicates that Sharia responsibility in digital transactions lies not only with the seller and buyer but also with the platform provider as the system designer. Algorithmic marketing mechanisms and the use of \textit{dark patterns} that encourage consumptive behavior and impulsive decisions not only have the potential to cause \textit{israf} (wastefulness/extravagance), but can also disrupt the user's sound reasoning, thus contradicting the primary objectives of Sharia (\textit{Maqashid al-Syariah}), especially the protection of wealth (\textit{Hifz al-Mal}) and the protection of intellect (\textit{Hifz al-'Aql}).

Based on all these findings, this research recommends the implementation of a comprehensive Sharia audit on \textit{e-commerce} platforms, which not only evaluates the validity of contracts but also ensures that algorithmic ethics, UI/UX design, and marketing patterns operate in alignment with Sharia values. This effort serves as a strategic step in realizing a digital economic ecosystem that is fair, transparent, and brings blessings to all its participants.

\bibliographystyle{./IEEEtran}
\bibliography{./IEEEabrv,./IEEEkelompok1}


\end{document}